\documentclass[letterpaper]{article} 
\usepackage{aaai2026}

\usepackage{times}  
\usepackage{helvet}  
\usepackage{courier}  
\usepackage[hyphens]{url}  
\usepackage{graphicx} 
\usepackage{natbib}  
\usepackage{caption} 
\usepackage{amsmath}
\usepackage{amssymb}
\usepackage{xcolor}
\usepackage[most]{tcolorbox}  

\definecolor{bg}{HTML}{d9d9d9}
\definecolor{text_bg}{HTML}{f9f9f9}
\tcbset{
  promptbox/.style={
    colback=white,            
    colframe=black,           
    coltitle=black,           
    colbacktitle=white,       
    fonttitle=\bfseries\fontsize{10}{14}\selectfont,   
    fontupper=\fontsize{9}{12}\selectfont,
    boxrule=0.4pt,            
    arc=1pt,                  
    left=6pt, right=6pt,
    top=4pt, bottom=4pt,      
    enhanced
  }
}

\usepackage{multirow} 
\usepackage{booktabs}
\usepackage{tabularx}

\usepackage{algorithm}
\usepackage{algorithmic}

\usepackage{newfloat}
\usepackage{listings}
\DeclareCaptionStyle{ruled}{labelfont=normalfont,labelsep=colon,strut=off} 
\floatstyle{ruled}
\newfloat{listing}{tb}{lst}{}
\floatname{listing}{Listing}
\title{Multimodal Recommendation via Self-Corrective Preference Alignment}
\author{
    Yalong Guan\textsuperscript{\rm 1}, 
    Xiang Chen\textsuperscript{\rm 1}, 
    Mingyang Wang\textsuperscript{\rm 2}, 
    Xiangyu Wu\textsuperscript{\rm 1}, 
    Lihao Liu\textsuperscript{\rm 1}, \\
    Chao Qi\textsuperscript{\rm 1}, 
    Shuang Yang\textsuperscript{\rm 1}, 
    Tingting Gao\textsuperscript{\rm 1}, 
    Guorui Zhou\textsuperscript{\rm 1}, 
    Changjian Chen\textsuperscript{\rm 2}\thanks{Corresponding Author}
}

\affiliations{
    \textsuperscript{\rm 1}Kuaishou Technology, Beijing, China \\
    \textsuperscript{\rm 2}
Hunan University of Computer Science and Electronic Engineering, Changsha, China
}

\usepackage{bibentry}

\begin{document}

\maketitle

\begin{abstract}

With the rapid growth of live streaming platforms, personalized recommendation systems have become pivotal in improving user experience and driving platform revenue.
The dynamic and multimodal nature of live streaming content (e.g., visual, audio, textual data) requires joint modeling of user behavior and multimodal features to capture evolving author characteristics.
However, traditional methods relying on single-modal features or treating multimodal ones as supplementary struggle to align users’ dynamic preferences with authors’ multimodal attributes, limiting accuracy and interpretability.
To address this, we propose MSPA(\textbf{M}ultimodal \textbf{S}elf-Corrective \textbf{P}reference \textbf{A}lignment), a personalized author recommendation framework with two components: (1) a Multimodal Preference Composer that uses MLLMs to generate structured preference text and embeddings from users’ tipping history; and (2) a Self-Corrective Preference Alignment Recommender that aligns these preferences with authors’ multimodal features to improve accuracy and interpretability.
Extensive experiments and visualizations show that MSPA significantly improves accuracy, recall, and text quality, outperforming baselines in dynamic live streaming scenarios.
\end{abstract}

\section{Introduction}

Personalized recommendation systems are vital for live streaming platforms, using historical user interactions, such as tipping and engagement records, to recommend relevant authors, thereby boosting user satisfaction and platform revenue. Unlike traditional recommendation systems, which rely on ID-based collaborative filtering with large-scale user-item interaction data, live streaming platforms like Kuaishou and TikTok Live have smaller user and author populations, rendering ID-based methods less effective due to sparse data. Content-based recommendation systems are thus essential, directly analyzing author content to capture user preferences. Live streaming content is inherently multimodal, combining visual signals (e.g., author appearance, scene setup), textual data (e.g., user comments, profiles), and audio features (e.g., tone, background music). Accurate recommendations require integrating these signals to capture authors’ dynamic traits. For instance, two authors may look similar but differ in content or atmosphere, such as interactive gaming versus calm vlogging, necessitating robust multimodal analysis. However, live streaming’s long-tail nature and lack of clear thematic labels pose significant challenges. Authors often switch topics or styles within a session or across time periods, such as from gaming to chatting, making static or single-modal methods, which focus on text or visuals alone, inadequate for capturing evolving content. This leads to a core difficulty: dynamically aligning user preferences with authors’ time-varying multimodal features, critical for precise recommendations in tipping-driven platforms.

However, existing methods struggle to capture the real-time, dynamic nature of live streaming content and fail to effectively integrate multimodal signals, posing a key challenge: dynamically aligning user preferences with the evolving multimodal features of authors. Some methods \cite{Intro_LLM_1, Intro_LLM_2, Intro_LLM_3, Intro_LLM_4, Intro_LLM_5} employ LLMs for sequence and preference modeling but largely ignore non-text modalities. Other methods \cite{Intro_MLLM_1, Intro_MLLM_2} utilize MLLMs to enhance feature representations for direct retrieval. However, they often fail to align user preferences with author content, as user interests tend to be abstract and lack structured explanatory text—resulting in weak semantic alignment and limited interpretability.

To address the challenges of multimodal integration, user-author alignment, and interpretability in live streaming recommendations, we propose the Multimodal Self-Corrective Preference Alignment (MSPA) framework. MSPA aligns user preferences with multimodal author features and generates structured explanatory text to enhance both accuracy and interpretability.
Our framework includes two components. First, we use a Multimodal Preference Composer that builds user representations by extracting visual, textual, audio-derived, and behavioral signals from past tipping behaviors. These signals are summarized into concise preference descriptions and embeddings that reflect user interests.
Second, we introduce a Self-Corrective Preference Alignment Recommender based on GRPO. It aligns user preferences with author features by sampling the model’s recommendations and comparing them to actual user feedback, such as tipping actions. This feedback serves as a natural supervision signal, allowing the model to refine its alignment strategy over time without relying on manually labeled data. As a result, the system can directly match users with suitable authors based on their preference representations, while also providing clear explanations rooted in those preferences.
Our contributions are as follows:
\begin{itemize}
    \item \textbf{GRPO-Based Self-Corrective Preference Alignment Recommender:} We leverage multimodal large language models to extract and integrate visual, textual, audio-derived, and behavioral signals from users’ tipping history, generating structured preference descriptions that provide a more accurate and interpretable representation of user interests.
    \item \textbf{MLLM-based Multimodal Preference Composer:} We propose a self-corrective recommendation module based on GRPO that samples model-generated matches and optimizes preference-author alignment using real user feedback, enabling dynamic adaptation without requiring human-labeled data and producing explanatory recommendation rationales.
    \item Extensive evaluations on benchmark datasets and visual analyses demonstrate that our method significantly improves recommendation accuracy, author recall, and explanation quality, effectively capturing dynamic user-author alignment in live streaming scenarios.
\end{itemize}

\section{Related Work}

\subsection{Traditional Recommendation Systems}
Traditional recommendation systems aimed to recommend items by inferring user interests based on their historical interactions with items and item features. 
Early recommendation systems typically relied on collaborative filtering \cite{traditional_rec_1, traditional_rec_2} and content-based filtering \cite{traditional_rec_3} to model user features. These methods focused on capturing statistical patterns from historical data, struggling to handle data sparsity and cold-start problems, where new users and items lacked sufficient interaction data. Furthermore, to balance computational efficiency and prediction accuracy, traditional recommendation systems typically employed a multi-stage cascaded architecture \cite{traditional_rec_4, traditional_rec_5, traditional_rec_6}. This architecture generally consisted of four stages:  retrieval, pre-ranking, ranking, and re-ranking. These methods typically treated each stage as an independent module, where the performance of one stage limited the effectiveness of the next, restricting the overall performance of the recommendation system. To address this, some existing methods \cite{traditional_rec_7, traditional_rec_8} improved recommendation performance by enhancing interactions between stages within the cascaded system.

\subsection{Integrating LLMs into Recommendation Systems}
The advancement of Large Language Models (LLMs) in Natural Language Processing has spurred growing interest in their application to recommendation systems.
Several studies leveraged LLMs as feature extractors to generate high-dimensional embeddings from textual or behavioral data, which were used to enhance traditional recommendation models \cite{LLM4Rec_1, LLM4Rec_2, LLM4Rec_3}.
Other works integrated LLMs to generate fine-grained token-level features, such as reasoning and factual knowledge, in order to enrich input representations and improve recommendation performance \cite{LLM4Rec_4, LLM4Rec_5}. 
Some methods directly used LLMs as the core recommendation model to generate item lists, supporting complex tasks such as multi-round or session-based recommendations \cite{LLM4Rec_6, LLM4Rec_7, LLM4Rec_8}.
Existing methods were typically limited to single-modal inputs, struggled to align user behavioral preferences with the multimodal features of authors, and often lacked interpretability. 
To address these limitations, we propose MSPA, a framework that integrates multimodal signals (visual, textual, audio) from live streaming content and employs self-corrective learning to align user preferences with authors’ dynamic attributes, improving recommendation accuracy and providing clear explanatory text for enhanced interpretability.

\section{Problem Formulation}

This study focuses on personalized author recommendation by generating user-specific behavioral preference texts from historical tipping sequences and recommending an author from the candidate set that best matches these preferences. It also provides structured explanations to enhance transparency and interpretability.

\subsection{Data Representation}

\noindent\textbf{Historical Tipping Sequence:} A user’s tipping history is represented as a sequence $ S = \{s_1, s_2, \dots, s_n\} $ of length $n$, where each author $ s_i $ is described by multimodal attributes $ (t_i, v_i, a_i, c_i) $, including:
\begin{itemize}
    \item Textual Profile ($ t_i $): the author’s personal introduction, such as nickname and location.
    \item Visual Features ($ v_i $): frames of live streaming.
    \item Audio-Derived Text ($ a_i $): transcripts of audio content.
    \item Comment Data ($ c_i $): user comments in the live room.
\end{itemize}

\noindent\textbf{Candidate authors:} For each user, a candidate set $ C = \{c_1, c_2, \dots, c_k\} $ is defined, containing $ k $ authors, each represented as $ c_j = (t_j, v_j, a_j, c_j) $. The set contains one ground-truth author $ c^* $ for evaluation.

\subsubsection{Preference Modeling}
Given a user’s historical tipping sequence $ S $, the goal is to generate a behavioral preference text $ P $ that summarizes the user’s interests, such as favored author types and live streaming environments. The generated text can be used to check user preferences and support downstream tasks like retrieval.

\subsubsection{Author Recommendation}
Given the user’s behavioral preference text $ P $ and a candidate set $ C $, the objective is to select the author $ \hat{c} \in C $ that best matches $ P $. The output is a pair $ (\hat{c}, e) $, where $ \hat{c} $ is the recommended author and $ e $ is a structured explanation detailing its alignment with $ P $. One ground-truth author $ c^* \in C $ is included for evaluation.

\section{Methodology}

\subsection{Overview}
\begin{figure*}[t!]
    \centering
    \includegraphics[width=\textwidth, height=\textheight, keepaspectratio]{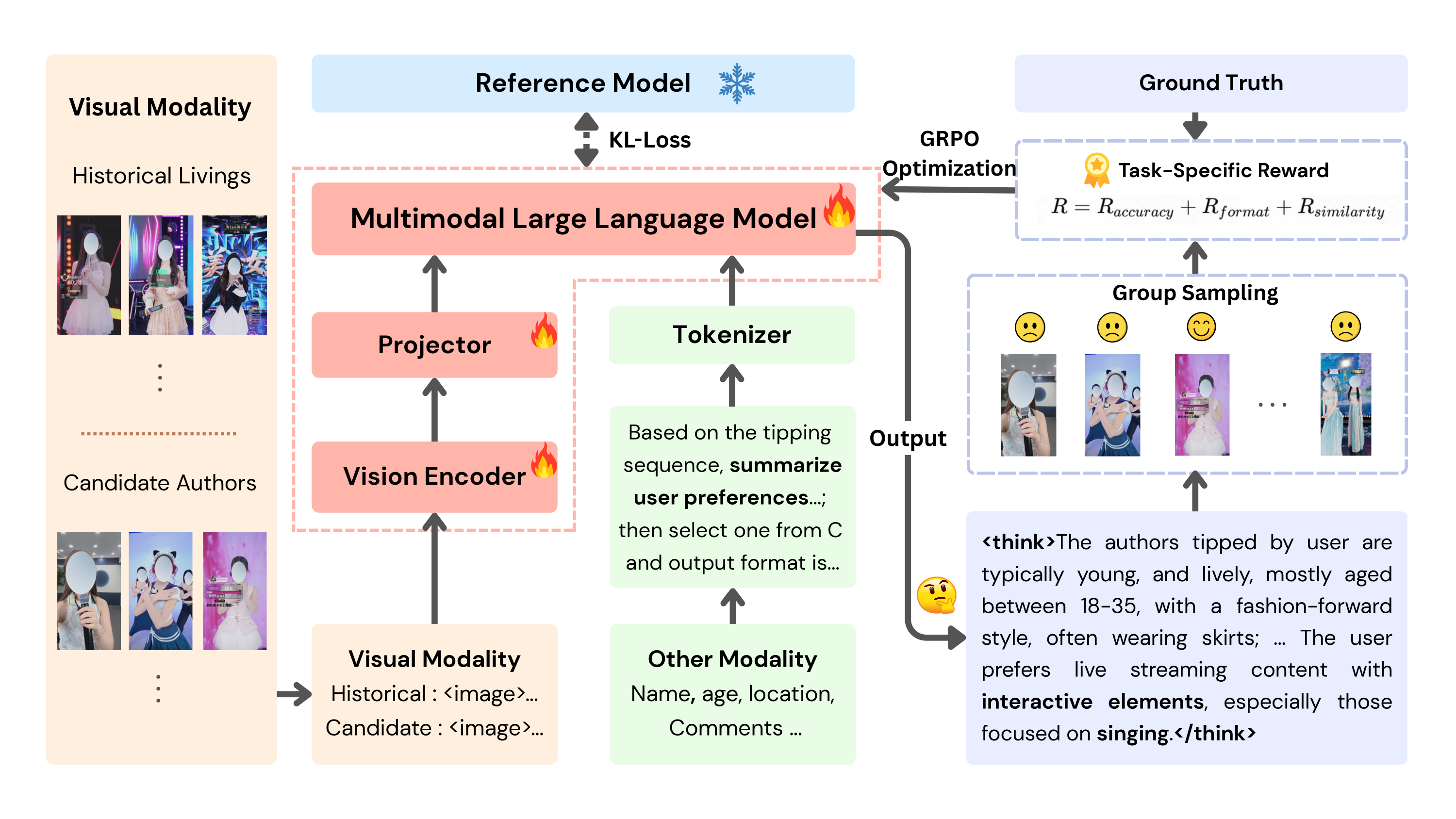}

    \caption{Overview of our framework. We propose the MSPA framework with two novel components to enhance author recommendation on live streaming platforms. The MLLM-based Multimodal Preference Composer conducts collaborative multimodal understanding to extract enhanced user preference features, while the GRPO-based Self-Corrective Preference Alignment Recommender simulates user behavior via deep reasoning. Leveraging GRPO-based reinforcement learning with behavior rewards, MSPA achieves superior alignment performance and significantly improves recommendation accuracy.}
    \label{fig:framework}
\end{figure*}


We propose MSPA, a unified framework for personalized author recommendation in live streaming platforms. As illustrated in Figure \ref{fig:framework}, MSPA consists of two key components. The first is the Multimodal Preference Composer, which utilizes MLLMs to extract and organize user behavioral preferences from historical tipping sequences by jointly modeling visual, textual, audio, and comment signals. The second is the Self-Corrective Preference Alignment Recommender, which employs GRPO-based reinforcement learning to simulate user behavior and align these preferences with candidate authors’ multimodal features. This novel framework separates preference modeling and alignment into two clear stages, forming the foundation of the overall method.


\subsection{MLLM-based Multimodal Preference Composer}
We introduce a multimodal preference modeling method based on MLLMs with parameters $\theta$, which takes a user’s historical tipping sequence $S$ as input and generates a behavioral preference text $P$ along with its embedding $P_{\text{emb}}$.

Our method models user behavioral preferences by understanding live streaming content:
\subsubsection{Multimodal Feature Extraction.} To model user behavioral preferences, we extract multimodal features from each tipped author $s_i \in S$. Specifically, we extract and encode the textual profile $t_i$, visual content $v_i$, audio-derived text $a_i$, and comment data $c_i$.

The textual inputs $t_i$, $a_i$, and $c_i$ are concatenated and tokenized using the MLLM tokenizer.
\[
T_i^{\text{text}} = \text{MLLM}_{\text{Tokenizer}}(t_i \oplus a_i \oplus c_i)
\]

The visual content $v_i$ is encoded using a Vision Transformer and aligned to the textual token embedding space of the MLLM using a projector:
\[
T_i^{\text{vis}} = \text{Projector}(\text{ViT}_{\theta}(v_i))
\]

The multimodal tokens for each $s_i$ are then concatenated:
\[
T_i = T_i^{\text{text}} \oplus T_i^{\text{vis}}
\]

All authors in the tipping sequence $S = \{s_1, s_2, \dots, s_n\}$ are processed in this way, and their token representations are concatenated to form:
\[
T_{\text{history}} = T_1 \oplus T_2 \oplus \dots \oplus T_n
\]

A structured natural language prompt is first tokenized into $T_{\text{prompt}}$, which guides the MLLM to extract the user's behavioral preferences from the multimodal features of the tipped authors in the history sequence. 
An example of extracting user behavioral preferences with the prompt is shown below:
\begin{tcolorbox}[title=Preference Extraction Prompt, promptbox]
\textbf{\texttt{<Instruction>:}} 
You are a professional analyst of user preferences for author recommendation. Based on the user's historical tipping sequence, analyze their behavioral preferences by summarizing multimodal signals from the previously tipped authors.
Please analyze the user's tipping preferences in terms of author types (e.g., talent, appearance...), live streaming environment (e.g., indoor/outdoor, lighting...), regional bias, and ..., and return a natural language description of the user's personalized behavioral preference. 

The user recently tipped the following authors. In tipping order, the multimodal features of each author are:

\textbf{\texttt{<input>:}} Images and texts of the tipping sequence \textit{S}.

\textbf{\texttt{<Answer Format>:}} ...
\end{tcolorbox}

The final input to the MLLM is constructed by concatenating $T_{\text{prompt}}$ with the sequence of multimodal tokens from the tipping history:
\[
T = T_{\text{prompt}} \oplus T_{\text{history}}
\]

\subsubsection{Preference Text Generation.}
The MLLM generates a behavioral preference text $P_{\text{text}}$ based on the input context $T$, summarizing the user's preferences over the multimodal features of authors in the historical tipping sequence:
\[
P_{\text{text}} = f_{\text{MLLM}}(T; \theta)
\]
For retrieval or similarity-based matching, the preference text \( P_{\text{text}} \) is encoded using a text encoder to extract the preference features:
\[
P_{\text{emb}} = \text{Encoder}(P_{\text{text}})
\]
Our method employs an MLLM to interpret multimodal live streaming content, capturing both the overall topic of each live streaming and subjective attributes such as atmosphere and author style, which are difficult to represent using predefined features, and thereby enhances the transparency and interpretability of the learned user preferences.

\subsection{GRPO-Based Self-Corrective Preference Alignment Recommender}

Given the preference text $P_{\text{text}}$ and a candidate set $C$, the model selects an author $\hat{c} \in C$ with a structured explanation $e$. The candidate set contains one ground-truth author $c^* \in C$, which serves as both a training supervision signal and an evaluation reference.

We propose a GRPO-based recommendation method that aligns user behavioral preferences with the multimodal features of candidate authors using MLLMs. The method employs a self-corrective reinforcement learning strategy that minimizes the discrepancy between the predicted author $\hat{c}$ and the ground-truth $c^*$. Unlike supervised fine-tuning (SFT) methods that rely on manually crafted responses, direct preference optimization (DPO) methods that require prompt--chosen--rejected triplets, and proximal policy optimization (PPO) methods that depend on training a reward model, GRPO optimizes the model using rule-based reward functions, offering a more stable and memory-efficient training strategy.

\subsubsection{Multimodal Recommendation.} Given the behavioral preference text $P_{\text{text}}$ and a candidate author set $C$, the objective is to prompt the MLLM to select the most relevant author $\hat{c} \in C$ and generate a structured explanation $e$ that aligns with the user’s preferences.

Each candidate $c_j \in C$ is encoded into a multimodal token sequence $T_j$, following the same process used in preference modeling.

A task-specific natural language prompt is tokenized into $T_{\text{prompt}}^{\text{rec}}$ and combined with the preference text $P_{\text{text}}$ to construct the input for the recommendation task.
\[
T^{\text{rec}} = T_{\text{prompt}}^{\text{rec}} \oplus \text{MLLM}_{\text{Tokenizer}}(P_{\text{text}}) \oplus T_1 \oplus T_2 \oplus \dots \oplus T_k
\]

The MLLM decoder processes $T^{\text{rec}}$ to generate a structured explanation $e$ in natural language, from which the recommended author $\hat{c}$ is extracted based on a predefined format:
\[
(\hat{c}, e) = f_{\text{MLLM}}(T^{\text{rec}}; \theta)
\]

An example of the recommendation prompt is shown below:

\begin{tcolorbox}[title=Author Recommendation Prompt, promptbox]
\textbf{\texttt{<Instruction>:}} Based on the user's behavioral preference above, choose the author the user is most likely to tip from the following options, explain the reasoning, including why the chosen author aligns with the user's preference and why others are not recommended.

\textbf{\texttt{<input>:}}

\textbf{\textit{A:}} Personal information: Names: \texttt{\{names\}}, Profiles: \texttt{\{descs\}},  Locations: \texttt{\{locations\}}...,

Live streaming screenshots: \texttt{<image>}...

\textbf{\textit{B:}} ...

\textbf{\texttt{<Answer Format>:}}

\{\texttt{"User Preference"}: ``...'', \texttt{"Recommendation Reason"}: ``...'', \texttt{"Answer"}: ``\textit{A/B/C/D...}''\}.
\end{tcolorbox}

\subsubsection{Self-Corrective Preference Alignment.}
To align the user's behavioral preferences with the multimodal features of candidate authors, we introduce a self-corrective learning framework based on Group Relative Policy Optimization (GRPO). This strategy allows the model to iteratively refine its predictions by comparing the recommended author $\hat{c}$ with the ground-truth label $c^*$, thereby optimizing its behavior without requiring human-labeled feedback.

Concretely, the predicted result and the ground-truth label are used to compute reward signals through a designed reward function, which guides the subsequent optimization of the model. This enables closed-loop self-corrective training and gradually improves alignment with user preferences.

\textbf{GRPO-Based Self-Corrective Optimization.}
In GRPO training, for each input query \( q \), a group of \( G \) outputs \( \{o_1, o_2, \dots, o_G\} \) is generated from the previous policy model \( \pi_{\theta_{\text{old}}} \), where \( G \) denotes the group size and \( o_i \) is the \( i \)-th output based on \( q \). GRPO optimizes the current policy \( \pi_\theta \) by leveraging the advantage function \( A_i \) to enhance output quality while ensuring stability. The advantage function is defined as:
\[
A_i = \frac{r_i - \text{mean}(\{r_1, r_2, \dots, r_G\})}{\text{std}(\{r_1, r_2, \dots, r_G\})},
\]
where \( r_i \) is the reward for output \( o_i \). To constrain divergence from a reference policy \( \pi_{\text{ref}} \), a KL divergence penalty is incorporated:
\[
\mathbb{D}_{\text{KL}}(\pi_\theta \| \pi_{\text{ref}}) = \frac{\pi_{\text{ref}}(o_i|q)}{\pi_\theta(o_i|q)} - \log \frac{\pi_{\text{ref}}(o_i|q)}{\pi_\theta(o_i|q)} - 1.
\]
GRPO balances optimization driven by \( A_i \) with the KL divergence constraint to improve \( \pi_\theta \) while maintaining stability.

\textbf{Reward Design}.
The reward serves as an optimization signal to align the user's behavioral preference representation $P_{\text{text}}$ with the multimodal features of candidates in $C$, allowing the model to map both into a shared semantic space and make accurate recommendations.

\begin{itemize}
    \item \textbf{Accuracy Reward:} 
    Given the predicted author $\hat{c}$ from the MLLM and the ground-truth author $c^*$, we define the accuracy reward as:
    \[
    R_{\text{accuracy}} =
    \begin{cases}
    1, & \text{if } \hat{c} = c^*; \\
    0, & \text{otherwise}.
    \end{cases}
    \]
    \item \textbf{Format Reward:}
    In addition to the accuracy reward, we incorporate a format reward that encourages the model to generate explanations in a structured manner. Specifically, the output $e$ must contain the key phrases ``User Preference'', ``Recommendation Reason'', and ``Recommended author''.
    \item \textbf{Similarity Reward:}
    Due to the scarcity of negative samples in the author recommendation task, using accuracy-based optimization can lead to unstable training. The model often recommends authors similar to the ground truth, leading to limited improvement, and direct penalization may cause over-correction. To address this, we introduce a similarity reward, which computes the cosine similarity between the predicted author $\hat{c}$ and the ground-truth author $c^*$ based on their multimodal features. The similarity is defined as:
    $$
    R_{\text{similarity}} = \text{sim}(\hat{c}, c^*).
    $$
    where $\text{sim}(\hat{c}, c^*)$ is the cosine similarity between the multimodal features of $\hat{c}$ and $c^*$, with values ranging from 0 to 1.
\end{itemize}

The final reward is defined as a weighted sum of the three components:
$$
R = \lambda_1 R_{\text{accuracy}} + \lambda_2 R_{\text{format}} + (1 - \lambda_1 - \lambda_2) R_{\text{similarity}}.
$$
where $\lambda_1 + \lambda_2 \leq 1$ balance the importance of the three reward components.

After GRPO-based training, the MLLM can directly recommend the most aligned author given a user’s behavioral preference text. Moreover, since the model has learned to align user preferences with author features, it can also be used independently to extract preference-aware representations of authors for retrieval or other downstream applications.

\section{Experiments}

\begin{table*}[!ht]
    \centering
    \small
    \begin{tabularx}{\textwidth}{ccccccc>{\centering\arraybackslash}X}
        \specialrule{0.12em}{0pt}{0pt}
        \rule{0pt}{12pt} \multirow{2}{*}{\textbf{Model}} & \multicolumn{6}{c}{\textbf{U2A}} & \textbf{A2A} \\ 
        \cline{2-8}
        \rule{0pt}{12pt} ~ & $\textbf{Acc}_{m=4}(\%)$ & $\textbf{Acc}_{m=10}(\%)$ & \textbf{Recall@5} & \textbf{NDCG@5} & \textbf{Recall@10} & \textbf{NDCG@10} & \textbf{A.R.} \\ 
        \hline
        \rule{0pt}{12pt} DeepSeek-VL2 & 15.32 & 8.71 & 0.031 & 0.027 & 0.052 & 0.033 & 0.461 \\
        \rule{0pt}{12pt} Qwen2.5-VL-7B & 47.00 & 24.24 & 0.052 & 0.037 & 0.115 & 0.057 & 0.558 \\ 
        \rule{0pt}{12pt} MiMo-VL-7B-RL & 66.93 & 55.94 & 0.115 & 0.081 & 0.177 & 0.100 & 0.760 \\ 
        \hline
        \rule{0pt}{12pt} \textbf{MSPA} & \textbf{77.78}(\textcolor{red}{\scriptsize +10.85}) & \textbf{66.67}(\textcolor{red}{\scriptsize +10.73}) & \textbf{0.250}(\textcolor{red}{\scriptsize+0.135}) & \textbf{0.170}(\textcolor{red}{\scriptsize +0.089}) & \textbf{0.281}(\textcolor{red}{\scriptsize +0.104}) & \textbf{0.179}(\textcolor{red}{\scriptsize +0.079}) & \textbf{0.809}(\textcolor{red}{\scriptsize +0.049}) \\
        \specialrule{0.12em}{0pt}{0pt}
    \end{tabularx}
    \caption{The performance of different methods.}
    \label{overall_performance}
\end{table*}

\subsection{Experimental Setup}
\subsubsection{Datasets.}We constructed a User to Author(U2A) dataset to validate the effectiveness of understanding and aligning user preferences and an Author to Author(A2A) dataset to measure the multi-modal feature representation capabilities of authors. Specifically, for the U2A dataset, we collected users' tipping behavior over the past 30 days as sessions for the recommendation task, with the last author tipped in each session as the
ground truth. During the data preprocessing, we filtered out users who tipped fewer than three authors or had insufficient tipping frequency, while also excluding authors lacking descriptive text or image information to ensure the quality of multimodal learning. Finally, we split the data into training, validation, and test sets with a 7:2:1 ratio, containing 7,000, 2,000, and 1,000 users, respectively.

For the A2A dataset, we manually constructed triples (author 1, author 2, author 3), where the similarity between author 1 and author 2 is greater than that between author 1 and author 3. By comparing the model’s and humans’ judgments of author similarity—computed from MLLM-extracted features—we assess how well the model aligns with human perception of author relationships.

\subsubsection{Baselines.}
We compare our method, MSPA, with several baseline models in the MLLM-based recommendation space. 1) DeepSeek-VL2 \cite{DeepSeek-VL2}, proposed by DeepSeek-AI company, is an advanced series of large Mixture-of-Experts (MoE) Vision-Language Models that significantly improve upon its predecessor, DeepSeek-VL. In this study, we use the 2.8B-parameter version of the model. 2) Qwen2.5-VL \cite{Qwen2.5-VL}, developed by Alibaba, is a multimodal large model that supports variable-length visual tokens and dynamic resolution, significantly enhancing visual performance, particularly when processing complex visual data. In this study, we use the 7B-parameter version of the model. 3) Mimo-VL-RL \cite{MiMo-VL-RL}, proposed by Xiaomi, is a 7B-parameter model that uses Reinforcement Learning with reasoning data to integrate multiple reward signals, significantly improving its visual reasoning and instruction-following abilities, thus enhancing recommendation effectiveness.

\subsubsection{Evaluation Metrics.}
For both U2A and A2A data, we first use MLLMs to generate the user’s behavioral preference text and the authors' feature texts. These texts are then encoded into feature vectors for further analysis.

The objective of the \textbf{U2A task} is to, given a user's tipping sequence, identify the author they are most likely to tip next. We use \textbf{$Acc_{m=4}$} and \textbf{$Acc_{m=10}$} to evaluate the model’s ability to select the ground-truth author from 4 and 10 candidates, respectively, measuring ranking precision. Additionally, Recall@K is used to evaluate whether the correct authors are successfully retrieved based on the user preferences and author features, while NDCG@K assesses the overall relevance and ranking quality of the retrieved results.

For the \textbf{A2A task}, we use the feature vectors to compute the similarity between author pairs in a triple (author 1, author 2, author 3), and the alignment is considered correct if the similarity between author 1 and author 2 is greater than that between author 1 and author 3. This allows us to compute the \textbf{alignment rate (A.R.)} to measure the consistency with the dataset.

These retrieval-based evaluations not only measure performance but also validate the effectiveness and explainability of the generated texts.

\subsubsection{Implementation Details.}
We use MiMo-VL-7B-RL as the MLLM backbone, trained on eight 80GB A800 GPUs. During training, we applied the GRPO \cite{GRPO} strategy for reinforcement learning, sampling 48 steps per iteration. We performed full-parameter fine-tuning with a learning rate set to 1e-6. For distributed training, we used the DeepSpeed \cite{DeepSpeed} framework and enabled ZeRO \cite{ZERO} Stage 3 to optimize memory usage and communication efficiency. To adapt to long text input scenarios, we set the maximum input length of the model to 12,800 tokens. The entire training process lasted approximately 7 days. In the recommendation process, MLLMs generate user preference descriptions and author text descriptions, which are then encoded by the BGE model \cite{BGE} for text feature extraction. We then perform recall tasks based on the extracted text features. Finally, we report on the performance of the module in the recommendation task. Due to the instability of the model output, our experimental results use the average of three generated results. All baseline inference parameters were aligned with those of our method.

\subsection{Overall Performance}
Table \ref{overall_performance} compares our method with baseline models under different scenarios. As shown, our method, MSPA, consistently outperforms all baselines in both the U2A and A2A tasks, achieving the highest accuracy and retrieval scores.

Specifically, in the U2A scenario, MSPA outperforms other models in both the accuracy of the given candidate selections and the recall of encoded features. This is due to our model's ability to effectively capture scene elements (such as furniture, author portraits, etc.) and demonstrate strong higher-level semantic understanding. For example, MSPA can summarize user preference signals, such as a preference for ``group or interactive live streaming content,'' and identify users with clear regional preferences (e.g., those who tend to watch local authors). These extracted features provide richer user preference information for subsequent recommendations.

In the A2A scenario, by modeling the textual features of authors, MSPA achieves the best alignment performance, demonstrating that the model's ability to distinguish between authors aligns more closely with human understanding. The model also shows a detailed understanding of live streaming content, including the appearance, attire, scene setup, and audience interactions of the authors, and it is capable of autonomously extracting relevant tags, which are crucial for evaluating the similarity between authors.

\begin{figure*}[t!]
    \centering
    \includegraphics[width=\textwidth, height=0.4\textheight, keepaspectratio]{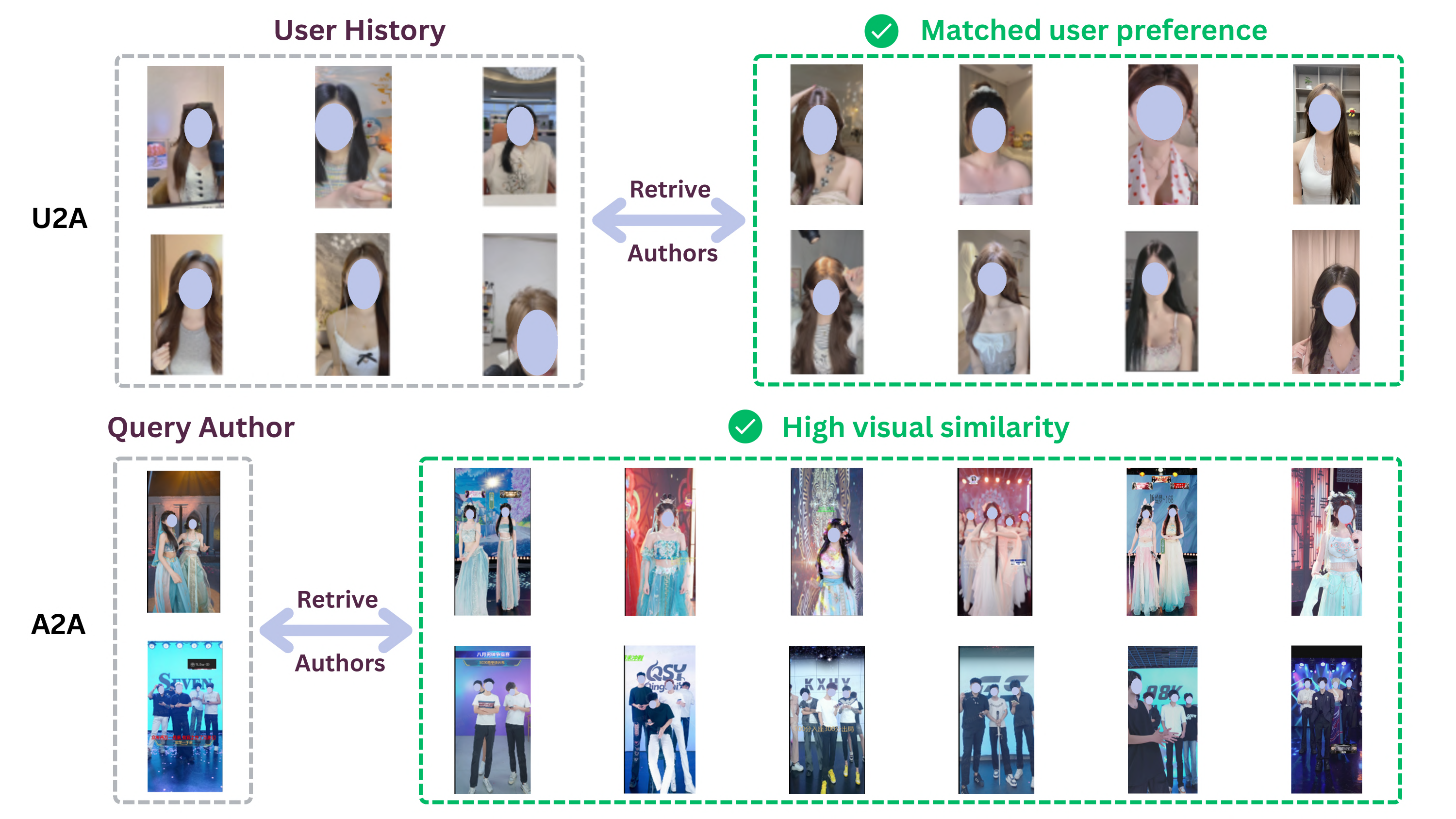}
    
    \caption{User-to-author and author-to-author retrieved author results.}
    \label{fig:U2A & A2A cases.}
\end{figure*}

\subsection{Offline Performance}
To evaluate the effectiveness of the proposed method in the recall stage, we replayed users' real online requests and calculated the Hit Rate@1000, which checks whether the live streaming viewed by the user appears in the top 1000 recommendations returned by the model. 

As a baseline, we used ID-based author embedding, while our method, MSPA, utilizes features derived from multimodal information after fine-tuning the MLLM, based on a behavioral preference text summarizing user preferences. According to the results in Table \ref{offline_results}, after integrating and aligning multimodal information, the user behavioral preferences and author multimodal features can better represent the relationship between users and authors, leading to improvements in both AUC and UAUC, measured in percentage points (\textbf{pp}). This demonstrates the effectiveness of the proposed method in real user requests.

\begin{table}[!ht]
    \centering
    \small
    \begin{tabularx}{\linewidth}{XXXXX}
        \specialrule{0.12em}{0pt}{0pt}
        \rule{0pt}{12pt} \multirow{2}{*}{\textbf{Model}} & \multicolumn{2}{c}{\textbf{Click}} & \multicolumn{2}{c}{\textbf{Gift}} \\
        \cline{2-5}
        \rule{0pt}{12pt} ~ & AUC & UAUC & AUC & UAUC \\
        \hline
        \rule{0pt}{12pt} Baseline & 0.8315 & 0.6416 & 0.9501 & 0.7167 \\
        \rule{0pt}{16pt} \textbf{MSPA} & 
        \begin{tabular}{@{}c@{}} \textbf{0.8321} \\ \textcolor{red}{\scriptsize (+0.06)} \end{tabular} & 
        \begin{tabular}{@{}c@{}} \textbf{0.6428} \\ \textcolor{red}{\scriptsize (+0.12)} \end{tabular} & 
        \begin{tabular}{@{}c@{}} \textbf{0.9513} \\ \textcolor{red}{\scriptsize (+0.12)} \end{tabular} & 
        \begin{tabular}{@{}c@{}} \textbf{0.7189} \\ \textcolor{red}{\scriptsize (+0.22)} \end{tabular} \\
        \specialrule{0.12em}{0pt}{0pt}
    \end{tabularx}
    \caption{Offline Results in term AUC of GAUC in Live-Streaming Rank Model.}
    \label{offline_results}
\end{table}

\subsection{User/Author-Author Visualization}
In this section, we demonstrate the retrieval performance of both the U2A and A2A tasks. We first encode the user’s behavioral preference text and the multimodal features of authors into vector representations, which are cached offline. Then, we perform an approximate nearest neighbor (ANN) search to retrieve the most relevant candidates. In the U2A task, the user embedding is used as the query to retrieve authors; in the A2A task, the embedding of a given author is used to retrieve other authors with similar styles. Figure \ref{fig:U2A & A2A cases.} shows one real example for each task, where the left side presents the query user (U2A) or author (A2A), and the right side displays the corresponding retrieval results.

From the Figure \ref{fig:U2A & A2A cases.}, we observe that MSPA not only retrieves candidates with semantically similar types (e.g., similar talents or genres), but also successfully captures visual details such as ``Hanfu'' elements in the scene, reflecting strong stylistic alignment. These results confirm that the representations learned by our method can effectively align users and authors within a shared semantic space, enabling high-quality personalized retrieval.

\subsection{Ablation Study}

\begin{table}[!ht]
    \centering
    \small
    \begin{tabularx}{\linewidth}{>{\centering\arraybackslash}Xcccc}
        \specialrule{0.12em}{0pt}{0pt}
        \rule{0pt}{12pt} \multirow{2}{*}{\textbf{Model}} & \multicolumn{3}{c}{\textbf{U2A}} & \textbf{A2A} \\
        \cline{2-5}
        \rule{0pt}{12pt} ~ & $\textbf{Acc}_{m=4}(\%)$ & \textbf{Recall@5} & \textbf{NDCG@5} & \textbf{A.R.} \\ 
        \hline
        \rule{0pt}{12pt} Baseline & 66.93 & 0.115 & 0.081 & 0.760 \\
        \hline
        \rule{0pt}{12pt} w/o ${R}_{similarity}$ & 73.08 & 0.208 & 0.122 & 0.791 \\ 
        \rule{0pt}{16pt} \textbf{MSPA} & 
        \begin{tabular}{@{}c@{}} \textbf{77.78} \\
        \textcolor{red}{\scriptsize (+4.7)} \end{tabular} &
        \begin{tabular}{@{}c@{}} \textbf{0.250} \\
        \textcolor{red}{\scriptsize (+0.042)} \end{tabular} &
        \begin{tabular}{@{}c@{}} \textbf{0.170} \\
        \textcolor{red}{\scriptsize (+0.048)} \end{tabular} &
        \begin{tabular}{@{}c@{}} \textbf{0.809} \\
        \textcolor{red}{\scriptsize (+0.018)} \end{tabular} \\
        \specialrule{0.12em}{0pt}{0pt}
    \end{tabularx}
    \caption{Ablation result.}
    \label{ablation_results}
\end{table}

To validate the effectiveness of our reward function design, we performed an ablation study comparing three models: the baseline MiMo-VL-7B-RL, the model trained without the Similarity Reward (\textbf{w/o ${R}_{similarity}$}), which uses Accuracy and Format Rewards, and the MSPA model with all reward functions.

The results in the first column of Table \ref{ablation_results} show that the model with Accuracy and Format Rewards improved the \textbf{$Acc_{m=4}$} metric from \textbf{66.93\% to 73.08\%}, demonstrating the benefit of Accuracy Reward in distinguishing authors. The full model, which includes all rewards, achieved the best overall performance of \textbf{77.78\%}, showing that the Similarity Reward enhances effectiveness. Additionally, we observed in our experiments that the model converged faster and exhibited more stable training behavior.

These findings demonstrate the effectiveness of our reward function design in improving recommendation accuracy and stability.

\section{Conclusion}

This paper presents MSPA, a personalized author recommendation framework that uses MLLMs to model user behavioral preferences from four-dimensional multimodal signals and employs Self-Corrective reinforcement learning to dynamically align user behavioral preferences with authors’ multimodal features, enhancing recommendation accuracy and interpretability.
In both U2A and A2A benchmark tests, as well as offline real user request replays, our method consistently outperforms baseline and ID-based feature models in recommendation accuracy and recall, demonstrating strong practical potential. Furthermore, visualization results intuitively confirm MSPA’s ability to align users and authors both semantically and stylistically.

\bibliography{aaai2026}

\begin{thebibliography}{30}
\providecommand{\natexlab}[1]{#1}

\bibitem[{Bai et~al.(2025)Bai, Chen, Liu, Wang, Ge, Song, Dang, Wang, Wang, Tang, Zhong, Zhu, Yang, Li, Wan, Wang, Ding, Fu, Xu, Ye, Zhang, Xie, Cheng, Zhang, Yang, Xu, and Lin}]{Qwen2.5-VL}
Bai, S.; Chen, K.; Liu, X.; Wang, J.; Ge, W.; Song, S.; Dang, K.; Wang, P.; Wang, S.; Tang, J.; Zhong, H.; Zhu, Y.; Yang, M.; Li, Z.; Wan, J.; Wang, P.; Ding, W.; Fu, Z.; Xu, Y.; Ye, J.; Zhang, X.; Xie, T.; Cheng, Z.; Zhang, H.; Yang, Z.; Xu, H.; and Lin, J. 2025.
\newblock Qwen2.5-VL Technical Report.
\newblock arXiv:2502.13923.

\bibitem[{Bao et~al.(2025)Bao, Zhang, Wang, Zhang, Yang, Luo, Chen, Feng, and Tian}]{LLM4Rec_7}
Bao, K.; Zhang, J.; Wang, W.; Zhang, Y.; Yang, Z.; Luo, Y.; Chen, C.; Feng, F.; and Tian, Q. 2025.
\newblock A Bi-Step Grounding Paradigm for Large Language Models in Recommendation Systems.
\newblock \emph{ACM Transactions on Recommender Systems}, 3(4): 1--27.

\bibitem[{Breese, Heckerman, and Kadie(2013)}]{traditional_rec_1}
Breese, J.~S.; Heckerman, D.; and Kadie, C. 2013.
\newblock Empirical Analysis of Predictive Algorithms for Collaborative Filtering.
\newblock arXiv:1301.7363.

\bibitem[{Chang et~al.(2023)Chang, Zhang, Fu, Zang, Guan, Lu, Hui, Leng, Niu, Song, and Gai}]{traditional_rec_5}
Chang, J.; Zhang, C.; Fu, Z.; Zang, X.; Guan, L.; Lu, J.; Hui, Y.; Leng, D.; Niu, Y.; Song, Y.; and Gai, K. 2023.
\newblock TWIN: TWo-stage Interest Network for Lifelong User Behavior Modeling in CTR Prediction at Kuaishou.
\newblock In \emph{Proceedings of the ACM SIGKDD International Conference on Knowledge Discovery and Data Mining}, 3785--3794.

\bibitem[{Chen et~al.(2024)Chen, Xiao, Zhang, Luo, Lian, and Liu}]{BGE}
Chen, J.; Xiao, S.; Zhang, P.; Luo, K.; Lian, D.; and Liu, Z. 2024.
\newblock M3-Embedding: Multi-Linguality, Multi-Functionality, Multi-Granularity Text Embeddings Through Self-Knowledge Distillation.
\newblock In \emph{Proceedings of the Findings of the Association for Computational Linguistics}, 2318--2335.

\bibitem[{Deng et~al.(2025)Deng, Wang, Cai, Ren, Hu, Ding, Luo, and Zhou}]{LLM4Rec_8}
Deng, J.; Wang, S.; Cai, K.; Ren, L.; Hu, Q.; Ding, W.; Luo, Q.; and Zhou, G. 2025.
\newblock OneRec: Unifying Retrieve and Rank with Generative Recommender and Iterative Preference Alignment.
\newblock arXiv:2502.18965.

\bibitem[{He et~al.(2017)He, Liao, Zhang, Nie, Hu, and Chua}]{traditional_rec_2}
He, X.; Liao, L.; Zhang, H.; Nie, L.; Hu, X.; and Chua, T.-S. 2017.
\newblock Neural Collaborative Filtering.
\newblock In \emph{Proceedings of the International Conference on World Wide Web}, 173--182.

\bibitem[{Hou et~al.(2022)Hou, Mu, Zhao, Li, Ding, and Wen}]{Intro_LLM_4}
Hou, Y.; Mu, S.; Zhao, W.~X.; Li, Y.; Ding, B.; and Wen, J.-R. 2022.
\newblock Towards Universal Sequence Representation Learning for Recommender Systems.
\newblock In \emph{Proceedings of the ACM SIGKDD International Conference on Knowledge Discovery and Data Mining}, 585--593.

\bibitem[{Huang et~al.(2023)Huang, Lian, Chen, Zheng, Xie, and Chen}]{traditional_rec_7}
Huang, X.; Lian, D.; Chen, J.; Zheng, L.; Xie, X.; and Chen, E. 2023.
\newblock Cooperative Retriever and Ranker in Deep Recommenders.
\newblock In \emph{Proceedings of the ACM Web Conference}, 1150--1161.

\bibitem[{Liu et~al.(2024{\natexlab{a}})Liu, Chen, Sakai, and Wu}]{LLM4Rec_4}
Liu, Q.; Chen, N.; Sakai, T.; and Wu, X.-M. 2024{\natexlab{a}}.
\newblock ONCE: Boosting Content-based Recommendation with Both Open- and Closed-source Large Language Models.
\newblock In \emph{Proceedings of the ACM International Conference on Web Search and Data Mining}, 452--461.

\bibitem[{Liu et~al.(2024{\natexlab{b}})Liu, Zheng, Huang, Li, Cai, Chai, Niu, Hui, Han, Mou, Wang, Bao, Yu, Zhou, Li, Song, Lian, and Gai}]{traditional_rec_6}
Liu, Q.; Zheng, K.; Huang, R.; Li, W.; Cai, K.; Chai, Y.; Niu, Y.; Hui, Y.; Han, B.; Mou, N.; Wang, H.; Bao, W.; Yu, Y.; Zhou, G.; Li, H.; Song, Y.; Lian, D.; and Gai, K. 2024{\natexlab{b}}.
\newblock RecFlow: An Industrial Full Flow Recommendation Dataset.
\newblock arXiv:2410.20868.

\bibitem[{Lops, de~Gemmis, and Semeraro(2011)}]{traditional_rec_3}
Lops, P.; de~Gemmis, M.; and Semeraro, G. 2011.
\newblock \emph{Content-based Recommender Systems: State of the Art and Trends}, 73--105.
\newblock Boston, MA: Springer US.

\bibitem[{Luo et~al.(2024)Luo, Cao, Sun, Yu, Huang, Yuan, Lin, Zheng, Wang, Hu, Qiu, Zhang, Zhang, Yan, Zhang, Zhang, Wen, Liu, Gai, and Zhou}]{Intro_MLLM_2}
Luo, X.; Cao, J.; Sun, T.; Yu, J.; Huang, R.; Yuan, W.; Lin, H.; Zheng, Y.; Wang, S.; Hu, Q.; Qiu, C.; Zhang, J.; Zhang, X.; Yan, Z.; Zhang, J.; Zhang, S.; Wen, M.; Liu, Z.; Gai, K.; and Zhou, G. 2024.
\newblock QARM: Quantitative Alignment Multi-Modal Recommendation at Kuaishou.
\newblock arXiv:2411.11739.

\bibitem[{Qiu et~al.(2021)Qiu, Wu, Gao, and Fan}]{LLM4Rec_2}
Qiu, Z.; Wu, X.; Gao, J.; and Fan, W. 2021.
\newblock U-BERT: Pre-training User Representations for Improved Recommendation.
\newblock \emph{Proceedings of the AAAI Conference on Artificial Intelligence}, 35(5): 4320--4327.

\bibitem[{Rajbhandari et~al.(2020)Rajbhandari, Rasley, Ruwase, and He}]{ZERO}
Rajbhandari, S.; Rasley, J.; Ruwase, O.; and He, Y. 2020.
\newblock ZeRO: Memory optimizations Toward Training Trillion Parameter Models.
\newblock In \emph{Proceedings of the International Conference for High Performance Computing, Networking, Storage and Analysis}, 1--16.

\bibitem[{Rasley et~al.(2020)Rasley, Rajbhandari, Ruwase, and He}]{DeepSpeed}
Rasley, J.; Rajbhandari, S.; Ruwase, O.; and He, Y. 2020.
\newblock DeepSpeed: System Optimizations Enable Training Deep Learning Models with Over 100 Billion Parameters.
\newblock In \emph{Proceedings of the ACM SIGKDD International Conference on Knowledge Discovery and Data Mining}, 3505--3506.

\bibitem[{Shao et~al.(2024)Shao, Wang, Zhu, Xu, Song, Bi, Zhang, Zhang, Li, Wu, and Guo}]{GRPO}
Shao, Z.; Wang, P.; Zhu, Q.; Xu, R.; Song, J.; Bi, X.; Zhang, H.; Zhang, M.; Li, Y.~K.; Wu, Y.; and Guo, D. 2024.
\newblock DeepSeekMath: Pushing the Limits of Mathematical Reasoning in Open Language Models.
\newblock arXiv:2402.03300.

\bibitem[{Sun et~al.(2019)Sun, Liu, Wu, Pei, Lin, Ou, and Jiang}]{Intro_LLM_3}
Sun, F.; Liu, J.; Wu, J.; Pei, C.; Lin, X.; Ou, W.; and Jiang, P. 2019.
\newblock BERT4Rec: Sequential Recommendation with Bidirectional Encoder Representations from Transformer.
\newblock In \emph{Proceedings of the ACM International Conference on Information and Knowledge Management}, 1441--1450.

\bibitem[{Wang and Lim(2023)}]{LLM4Rec_6}
Wang, L.; and Lim, E.-P. 2023.
\newblock Zero-Shot Next-Item Recommendation using Large Pretrained Language Models.
\newblock arXiv:2304.03153.

\bibitem[{Wang, Lin, and Metzler(2011)}]{traditional_rec_4}
Wang, L.; Lin, J.; and Metzler, D. 2011.
\newblock A cascade ranking model for efficient ranked retrieval.
\newblock In \emph{Proceedings of the International ACM SIGIR Conference on Research and Development in Information Retrieval}, 105--114.

\bibitem[{Wang et~al.(2024)Wang, Wang, Yang, Wen, Kong, Li, and Gai}]{traditional_rec_8}
Wang, Y.; Wang, Z.; Yang, J.; Wen, S.; Kong, D.; Li, H.; and Gai, K. 2024.
\newblock Adaptive Neural Ranking Framework: Toward Maximized Business Goal for Cascade Ranking Systems.
\newblock In \emph{Proceedings of the ACM Web Conference}, 3798--3809.

\bibitem[{Wu et~al.(2021{\natexlab{a}})Wu, Wu, Qi, and Huang}]{LLM4Rec_1}
Wu, C.; Wu, F.; Qi, T.; and Huang, Y. 2021{\natexlab{a}}.
\newblock Empowering News Recommendation with Pre-trained Language Models.
\newblock In \emph{Proceedings of the International ACM SIGIR Conference on Research and Development in Information Retrieval}, 1652--1656.

\bibitem[{Wu et~al.(2021{\natexlab{b}})Wu, Wu, Yu, Qi, Huang, and Xie}]{LLM4Rec_3}
Wu, C.; Wu, F.; Yu, Y.; Qi, T.; Huang, Y.; and Xie, X. 2021{\natexlab{b}}.
\newblock UserBERT: Contrastive User Model Pre-training.
\newblock arXiv:2109.01274.

\bibitem[{Wu et~al.(2022)Wu, Magnani, Chaidaroon, Puthenputhussery, Liao, and Fang}]{Intro_LLM_2}
Wu, X.; Magnani, A.; Chaidaroon, S.; Puthenputhussery, A.; Liao, C.; and Fang, Y. 2022.
\newblock A Multi-task Learning Framework for Product Ranking with BERT.
\newblock In \emph{Proceedings of the ACM Web Conference}, 493--501.

\bibitem[{Wu et~al.(2024)Wu, Chen, Pan, Liu, Liu, Dai, Gao, Ma, Wu, Wang, Xie, Wu, Hu, Wang, Sun, Li, Piao, Guan, Liu, Xie, You, Dong, Yu, Zhang, Zhao, Wang, and Ruan}]{DeepSeek-VL2}
Wu, Z.; Chen, X.; Pan, Z.; Liu, X.; Liu, W.; Dai, D.; Gao, H.; Ma, Y.; Wu, C.; Wang, B.; Xie, Z.; Wu, Y.; Hu, K.; Wang, J.; Sun, Y.; Li, Y.; Piao, Y.; Guan, K.; Liu, A.; Xie, X.; You, Y.; Dong, K.; Yu, X.; Zhang, H.; Zhao, L.; Wang, Y.; and Ruan, C. 2024.
\newblock DeepSeek-VL2: Mixture-of-Experts Vision-Language Models for Advanced Multimodal Understanding.
\newblock arXiv:2412.10302.

\bibitem[{Xi et~al.(2024)Xi, Liu, Lin, Cai, Zhu, Zhu, Chen, Tang, Zhang, and Yu}]{LLM4Rec_5}
Xi, Y.; Liu, W.; Lin, J.; Cai, X.; Zhu, H.; Zhu, J.; Chen, B.; Tang, R.; Zhang, W.; and Yu, Y. 2024.
\newblock Towards Open-World Recommendation with Knowledge Augmentation from Large Language Models.
\newblock In \emph{Proceedings of the ACM Conference on Recommender Systems}, 12--22.

\bibitem[{Xiaomi et~al.(2025)Xiaomi, :, Xia, Shen, Cici, Zhu, Zhang, Wang, Zhang, Liu, Xiao, Dong, Zhao, Li, Wang, Yu, Chen, Wang, Ma, Deng, Huang, Song, Jiang, Ye, Cai, He, Zhang, Zhang, Wang, Tian, Zhao, Qu, Xu, Shi, Bao, Fang, Zhou, Zhou, Li, Zhu, Chen, Wang, Liu, Li, Gu, Ren, Liu, Deng, Zhuang, Lv, Yang, Zhang, Yong, Zhang, Song, Xu, Wang, Yan, Tu, Tian, Wang, Yu, Lin, Song, and Yue}]{MiMo-VL-RL}
Xiaomi, L.-C.; :; Xia, B.; Shen, B.; Cici; Zhu, D.; Zhang, D.; Wang, G.; Zhang, H.; Liu, H.; Xiao, J.; Dong, J.; Zhao, L.; Li, P.; Wang, P.; Yu, S.; Chen, S.; Wang, W.; Ma, W.; Deng, X.; Huang, Y.; Song, Y.; Jiang, Z.; Ye, B.; Cai, C.; He, C.; Zhang, D.; Zhang, D.; Wang, G.; Tian, H.; Zhao, H.; Qu, H.; Xu, H.; Shi, J.; Bao, K.; Fang, K.; Zhou, K.; Zhou, K.; Li, L.; Zhu, M.; Chen, N.; Wang, Q.; Liu, S.; Li, S.; Gu, S.; Ren, S.; Liu, S.; Deng, S.; Zhuang, W.; Lv, W.; Yang, W.; Zhang, X.; Yong, X.; Zhang, X.; Song, X.; Xu, X.; Wang, X.; Yan, Y.; Tu, Y.; Tian, Y.; Wang, Y.; Yu, Y.; Lin, Z.; Song, Z.; and Yue, Z. 2025.
\newblock MiMo: Unlocking the Reasoning Potential of Language Model -- From Pretraining to Posttraining.
\newblock arXiv:2505.07608.

\bibitem[{Yang et~al.(2023)Yang, Chen, Jiang, Cho, Huang, and Lu}]{Intro_LLM_5}
Yang, F.; Chen, Z.; Jiang, Z.; Cho, E.; Huang, X.; and Lu, Y. 2023.
\newblock PALR: Personalization Aware LLMs for Recommendation.
\newblock arXiv:2305.07622.

\bibitem[{Yang et~al.(2022)Yang, Qiao, Shao, Yan, and Yang}]{Intro_LLM_1}
Yang, Y.; Qiao, Y.; Shao, J.; Yan, X.; and Yang, T. 2022.
\newblock Lightweight Composite Re-Ranking for Efficient Keyword Search with BERT.
\newblock In \emph{Proceedings of the ACM International Conference on Web Search and Data Mining}, 1234--1244.

\bibitem[{Zhang et~al.(2025)Zhang, Zhang, Wu, Wu, Xu, Zhao, Gao, Hu, and Chen}]{Intro_MLLM_1}
Zhang, C.; Zhang, H.; Wu, S.; Wu, D.; Xu, T.; Zhao, X.; Gao, Y.; Hu, Y.; and Chen, E. 2025.
\newblock NoteLLM-2: Multimodal Large Representation Models for Recommendation.
\newblock In \emph{Proceedings of the ACM SIGKDD International Conference on Knowledge Discovery and Data Mining}, 2815--2826.

\end{thebibliography}

\end{document}